\RequirePackage{fix-cm}
\documentclass[smallextended]{svjour3}       % onecolumn (second format)
\smartqed  % flush right qed marks, e.g. at end of proof
\usepackage{graphicx}
\usepackage{natbib}
\usepackage{graphics}
\usepackage{url}

\usepackage{amsmath,amssymb}
 
\usepackage[table]{xcolor}
\usepackage{lscape}
\usepackage{ulem}

\renewcommand{\arraystretch}{2.5}
 
\definecolor{verdone}{rgb}{0,0.398,0}
\definecolor{stateblue}{rgb}{0.14, 0.0937, 0.508}
\definecolor{arancio}{rgb}{1,0.398,0}
\definecolor{rosso}{rgb}{0.8,0,0}
\definecolor{grigioc}{rgb}{0.4,0.4,0.4}
\definecolor{violet}{rgb}{0.7,0,0.6}
\definecolor{azzurro}{rgb}{0.05, 0.45, 0.75}
\definecolor{marrone}{rgb}{0.28, 0.04, 0.00}
\definecolor{bold}{cmyk}{0,1.0,1.0,0.30}
\newcommand{\deltanu}{\mbox{$\Delta\nu$}}
\newcommand{\numax}{\mbox{$\nu_\mathrm{max}$}}
\newcommand{\hideit}[1]{}

\begin{document}

\title{HAYDN} %\thanks{Grants or other notes

\subtitle{High-precision AsteroseismologY of DeNse stellar fields}


\author{ Andrea Miglio 
\and L\'eo Girardi 
\and Frank Grundahl
\and Benoit Mosser 
\and Nate Bastian 
\and Angela Bragaglia
\and Karsten Brogaard 
\and Ga\"el Buldgen
\and William Chantereau
\and William Chaplin 
\and Cristina Chiappini
\and Marc-Antoine Dupret
\and Patrick Eggenberger
\and Mark Gieles
\and Robert Izzard 
\and Daisuke Kawata
\and Christoffer Karoff
\and Nad\`ege Lagarde
\and Ted Mackereth
\and Demetrio Magrin
\and Georges Meynet
\and Eric Michel
\and Josefina Montalb\'an
\and Valerio Nascimbeni
\and Arlette Noels
\and Giampaolo Piotto
\and Roberto Ragazzoni
\and Igor Soszy{\'n}ski
\and Eline Tolstoy
\and Silvia Toonen
\and Amaury Triaud
\and Fiorenzo Vincenzo\thanks{A list of people who have expressed interest for HAYDN is available here:\\ {\texttt{https://www.asterochronometry.eu/haydn/people.html}}.}}

\institute{Andrea Miglio \at
School of Physics and Astronomy, University of Birmingham, UK
              \email{andrea.miglio@unibo.it}           %  \\
\at{Dipartimento di Fisica e Astronomia, Universit{\`a} degli Studi di Bologna, Bologna, Italy}
\at{INAF-OAS Bologna, Italy}
\and L\'eo Girardi \at {INAF Osservatorio Astronomico di Padova, Italy}
\and Frank Grundahl  \at {Stellar Astrophysics Centre (SAC), Aarhus, Denmark}
\and Benoit Mosser  \at {LESIA, Observatoire de Paris, Universit\'e PSL, CNRS, Sorbonne Universit\'e, Universit\'e de Paris, 92195 Meudon, France}
\and Nate Bastian  \at {ARI, Liverpool John Moores University, Liverpool, UK}
\and Angela Bragaglia  \at {INAF-OAS Bologna, Italy}
\and Karsten Brogaard  \at {Stellar Astrophysics Centre (SAC), Aarhus, Denmark} 
\at{Astronomical Observatory, Institute of Theoretical Physics and Astronomy, Vilnius University, 10257 Vilnius, Lithuania}
\and Ga\"el Buldgen  \at {Observatoire de Gen\`eve, Universit\'e de Gen\`eve, Switzerland} 
\and William Chantereau  \at {ARI, Liverpool John Moores University, Liverpool, UK}
\and William Chaplin  \at {School of Physics and Astronomy, University of Birmingham, UK}
\and Cristina Chiappini  \at {AIP, Potsdam, Germany}
\and Marc-Antoine Dupret  \at {STAR Institute, University of Li\`ege, Belgium}
\and Patrick Eggenberger  \at {Observatoire de Gen\`eve, Universit\'e de Gen\`eve, Switzerland}
\and Mark Gieles  
\at {ICREA, Pg. Llu\'{i}s Companys 23, 08010 Barcelona, Spain}
\at {Institut de Ciències del Cosmos (ICCUB), Universitat de Barcelona (IEEC-UB), E08028 Barcelona, Spain }
%\at {ICREA \& ICCUB, Universitat de Barcelona, Spain}
\and Robert Izzard  \at {Astrophysics Research Group, University of Surrey, Guildford, UK}
\and Daisuke Kawata  \at {MSSL, University College London, Surrey, UK}
\and Christoffer Karoff  \at {Stellar Astrophysics Centre (SAC), Aarhus, Denmark}
\and Nad\`ege Lagarde  \at {Institut UTINAM, UMR6213, OSU THETA, Observatoire de Besan\c con, France}
\and Ted Mackereth  \at {School of Physics and Astronomy, University of Birmingham, UK}
\at{Canadian Institute for Theoretical Astrophysics, University of Toronto, Toronto, Canada}
\at{Dunlap Institute for Astronomy and Astrophysics, University of Toronto, Toronto, Canada}
\at {David A. Dunlap Department for Astronomy and Astrophysics, University of Toronto,  Toronto, Canada }
\and Demetrio Magrin  \at {INAF Osservatorio Astronomico di Padova, Italy}
\and Georges Meynet  \at {Observatoire de Gen\`eve, Universit\'e de Gen\`eve, Switzerland} 
\and Eric Michel  \at {LESIA, Observatoire de Paris, Meudon, France}
\and Josefina Montalb\'an  \at {School of Physics and Astronomy, University of Birmingham, UK}
\and Valerio Nascimbeni  \at {INAF  Osservatorio Astronomico di Padova, Italy}
\and Arlette Noels  \at {STAR Institute, University of Li\`ege, Belgium}
\and Giampaolo Piotto  \at {Dipartimento di Fisica e Astronomia, Universit\`a di Padova, Italy}
\and Roberto Ragazzoni  \at {INAF Osservatorio Astronomico di Padova, Italy}
\and Igor Soszy{\'n}ski  \at {Astronomical Observatory, University of Warsaw, Poland}
\and Eline Tolstoy  \at {Kapteyn Astronomical Institute, University of Groningen, NL}
\and Silvia Toonen  \at {School of Physics and Astronomy, University of Birmingham, UK}
\at{Anton Pannekoek Institute, University of Amsterdam, Amsterdam, the
Netherlands}
\and Amaury Triaud  \at {School of Physics and Astronomy, University of Birmingham, UK}
\and Fiorenzo Vincenzo  \at {School of Physics and Astronomy, University of Birmingham, UK}
\at {Department of Astronomy \& Center for Cosmology and AstroParticle Physics, The Ohio State University, Columbus, USA }
}
\date{Received: date / Accepted: date}
\maketitle
\begin{abstract}
In the last decade, the \textit{Kepler} and CoRoT space-photometry missions have demonstrated the potential of asteroseismology as a novel, versatile and powerful tool to perform exquisite tests of stellar physics, and to enable precise and accurate characterisations of stellar properties, with impact on both exoplanetary and Galactic astrophysics. 
Based on our improved understanding of the strengths and limitations of such a tool, we argue for a new small/medium space mission dedicated to gathering high-precision, high-cadence, long photometric series in dense stellar fields. 
 Such a mission will lead to breakthroughs in stellar astrophysics, especially in the metal poor regime, will elucidate the evolution and  formation of open and globular clusters, and aid our understanding 
of the assembly history and chemodynamics of the Milky Way's bulge and a few nearby dwarf galaxies.

\end{abstract}
\section{Context and motivation}
Primarily driven and motivated by the impetus behind the search for transiting exoplanets, high-precision, high-cadence, long photometric surveys from space have also led to major advances in the study and interpretation of global, resonant pulsation modes in stars (asteroseismology).

The discoveries enabled by asteroseismic studies with data from CoRoT, \textit{Kepler}, K2, and now TESS,  not only are revolutionising our knowledge of stellar evolution, but have also demonstrated the relevance and potential of this novel technique well beyond the realm of stellar physics. The key role of asteroseismology in providing precise and accurate characterisations of planet-host stars is  widely recognised and, e.g., is deeply rooted into the core programme of the future ESA PLATO mission.
Moreover, asteroseismic constraints for ensembles of stars (in particular red giants) have given us the ability to measure masses (hence infer ages) of tens of thousands of stars across the Galaxy.  Coupling asteroseismic constraints to precise and accurate astrometric data from Gaia, and data from large-scale spectroscopic surveys, enables the study of the chemical and dynamical evolution of our Galaxy with unprecedented temporal resolution. 

To some extent, results from the exploitation of asteroseismic data went even beyond the expectations of those who, starting from the early 1980s, have been advocating for a European space mission dedicated to stellar seismology \citep[see e.g.][for a review on the formidable foresight and decades-long efforts leading eventually to space-based asteroseismology]{Roxburgh2002}. 
As discussed in more detail below, results from \textit{Kepler} and CoRoT have primarily demonstrated the potential of asteroseismology as a versatile and powerful tool. With the benefit of hindsight, i.e. having now a better understanding of the strengths and limitations of such a tool, we can propose a mission design that will lead to breakthroughs in three broad areas:
\begin{enumerate}
    \item high-precision stellar astrophysics, especially in the metal poor regime,
    \item evolution and formation of stellar clusters, 
    \item assembly history and chemical evolution of the Milky Way's bulge and a few nearby dwarf galaxies.
\end{enumerate}

We argue that a mission dedicated to high-precision, high-cadence, long photometric time series to dense stellar fields is needed to address specific, long-standing key questions in those three areas, which we describe in detail in Sec. \ref{sec:obj}. 

For instance,  high-quality asteroseismic constraints for large samples of coeval and initially-chemically-homogeneous stars in open and globular clusters, and in different evolutionary stages, will then be coupled with spectroscopic and astrometric data to assemble high-precision stellar physics laboratories to test stellar evolution models to unprecedented precision. 

The paper is organised as follows: in Sec. \ref{sec:obj} we describe in detail the proposed science questions. We then translate scientific requirements into a list of targets and we provide a possible mission design aided by simulations / calibrations using \textit{Kepler} data (Sec. \ref{sec:targets}). 
Before delving into the science objectives, we provide  below a succinct summary about the inferences on stellar structure that one can make from the observation of solar-like oscillations (Sec.~\ref{sec:preamble}), and clarify the limitations of present and planned asteroseismic missions (Sec.~\ref{sec:limits}).

\subsection{Preamble - what can we infer from high-precision, high-cadence, long photometric time series of solar-like oscillators?} 
\label{sec:preamble}
Among the various classes of pulsating stars, we focus here primarily on those showing rich spectra of solar-like oscillations, which are global oscillation modes excited and intrinsically damped by turbulence in the outermost layers of convective envelopes. 

These oscillations are ubiquitous in stars with sufficiently deep convective envelopes, hence they can be detected in low- and intermediate-mass stars belonging to different evolutionary stages (main sequence, subgiant phase, along the evolution on the red-giant branch, in the core-He burning phase and the early {asymptotic giant branch (AGB)}). This means that we can explore large mass (approximately 0.8-3~M$_\odot$) and age intervals, including the low-mass, low-metallicity domain, which is key to studies of the early Universe.
At the same time, e.g. within a given cluster, we can study the structures and properties of a wide range of coeval stars, using a homogeneous sample. 

Moreover, solar-like oscillations show very rich, yet relatively simple frequency patterns that one can robustly interpret aided by well-established stellar pulsation theory. The information on global and local features of stellar interiors that can be extracted from such spectra is succinctly presented below \citep[see][for recent, comprehensive, reviews on the subject]{Chaplin2013, Hekker2017, Garcia2019}.

\begin{figure}
\centering
  \includegraphics[width=\linewidth]{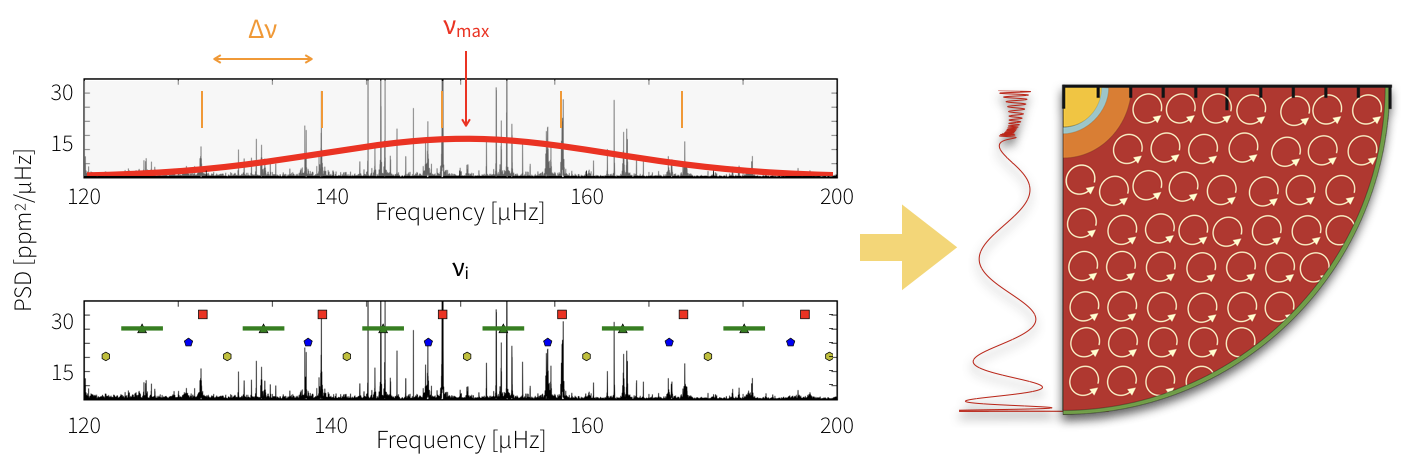}
\caption{Power spectra based on \textit{Kepler}'s photometric observations of a red-giant stars (KIC12008916). The \textit{upper-left panel} shows examples of the average seismic parameters large frequency separation (\deltanu) between radial modes and have illustrated the oscillation envelope of power (red thick line), the maximum of which is labelled as \numax. The \textit{lower-left} panel shows the complex, rich spectrum of individual mode frequencies, which can be however robustly interpreted and used to make detailed inferences about both global and local stellar features (see Sec. \ref{sec:preamble}). Symbols represent the radial modes (red squares), the dipole mixed pressure-gravity modes (green triangles with extended bars), quadrupole modes (blue pentagons), and octupole modes (yellow hexagons). The \textit{right panel} shows a cartoon of the internal structure of a red-giant-branch star, together with an example of the perturbation associated to the radial displacement of a mixed pressure-gravity mode, which is able to probe both the envelope and the core of such a star.}
\label{fig:PS}
\end{figure}

\begin{itemize}
    \item Inference on \textit{global stellar properties}. High-precision mean densities, masses, radii, ages, and envelope He mass fraction can be inferred via detailed forward and inverse modelling techniques. This can be done using asteroseismic constraints of different quality: from average seismic parameters to individual-mode frequencies (see Fig. \ref{fig:PS}). Moreover, seismically inferred radii coupled with effective temperature allow inferring distances, with typical uncertainties of a few percent, which make them competitive with Gaia for the faint targets.
    \item Constraints on the \textit{internal rotation rate} of stars. These are possible once rotational frequency splittings are measured, which is generally possible both in { main-sequence (MS)} stars (rotational splitting of pressure (p) modes, probing the envelope/surface rotation) and in subgiants and giants (core rotation inferred from the splitting of gravity (g) and mixed modes, which we do observe in evolved stars). In evolved stars (e.g., stars showing mixed p-g modes) it is primarily the core rotation we can measure, with perhaps subtleties on more detailed, local inferences on the rotational profile \citep{Beck2012, DiMauro2016, Deheuvels2012, Mosser2013, Gehan2018}. In MS stars, it is the envelope rotation which is accessible to observations, including information on latitudinal differential rotation \citep{Lund2014, Nielsen2017, Bazot2019, Benomar2018}.
    \item  The \textit{inclination angles of the star's spin axis} with respect to the line of sight can potentially be inferred from measuring the relative amplitudes of the different components of rotationally split frequencies \citep[e.g., ][]{Gizon2003, Chaplin2013b, Huber2013, corsaro17, Kuszlewicz2019}.
    \item Detailed constraints on the \textit{internal structure} of stars can be inferred by looking at specific signatures in the oscillation frequencies. To a first approximation, from the (mixed) oscillation modes patterns one can infer average quantities characteristics of high-order p and g modes, i.e., respectively, large frequency spacing (\deltanu) and period spacing ($\Delta P$).  \deltanu\, (p modes) is primarily a proxy for the stellar mean density, while $\Delta P$ is determined by the thermal and chemical stratification of the radiative, near-core regions. The latter can be used to e.g. infer the mass of the inert helium core on the red giant branch (RGB), and the detailed density/chemical composition profiles of the inner regions in core-He burning stars 
        \citep[e.g.][]{Montalban2013, bossini15, Constantino2015, Cunha2015}. Moreover from the observational signature of the coupling between the pressure- and gravity-modes cavity, one can infer  very local properties of the evanescent region between the two coupled propagation cavities \citep{Takata2016, Mosser2017b}.
     \item Deviations from the expected, approximated, frequency patterns of p and g modes can be used to detect and infer detailed properties of \textit{localised gradients} in e.g. the chemical composition / thermal stratification / sound speed profile (referred to as ``glitches''). 
     Examples of such localised features include convective boundaries (regions of strong thermal and chemical gradients) and He-ionisation region where the adiabatic index, hence sound speed, has a local minimum, and whose signature can be used to infer the envelope He abundance \citep[e.g.][]{Vorontsov1988, Monteiro1994, Miglio2010,Mazumdar2014, Broomhall2014,  Verma2019, mckeever19}.
     \item Determination of model independent properties through \textit{seismic inversions} are possible  with high-quality (i.e. \textit{Kepler}-like) data.   
     These approaches can be used to robustly infer global properties \citep[e.g. mean density, see][]{Reese2012, Buldgen2019}, but also to build specific proxies \citep{Roxburgh2002a, Buldgen2015a, Buldgen2015b, Buldgen2018} of stellar structure designed to test key ingredients of stellar physics related to,  e.g., microscopic diffusion, convective boundary mixing (see Sec. \ref{sec:SO1}).     \item The signature of \textit{activity cycles} can be found in frequency shifts / modes' lifetimes changes as a function of time \citep[e.g., ][]{Garcia2010, Kiefer2017, Santos2018, Salabert2018}. In some cases a coarse latitudinal dependence of the magnetic field may be inferred from the angular dependence of the visibility of the rotationally split modes \citep{Karoff2018,Thomas2019}. 
\end{itemize}

The precision and robustness with which one can achieve such detailed constraints on stars is strongly dependent on the length and quality of the photometric time-series available. The analysis requires, for instance, long-duration time series to obtain the requisite frequency resolution for extracting clear signatures of rotation in the oscillation spectrum. Examples of how the different duration of a dataset affects the seismic inferences is presented e.g. in \citet{DaviesMiglio2016} and, extensively, in \citet{Mosser2019}. 

The mapping from the science questions outlined in Sec. \ref{sec:obj} into observational tests and, finally, into a possible mission design (Sec.  \ref{sec:targets}) will in fact be greatly aided by the knowledge acquired on the analysis and interpretation of existing seismic data.

\subsection{Limitations of present/planned asteroseismic missions}
\label{sec:limits}

In spite of being mainly planet-hunting missions,  CoRoT \citep{Baglin2009} and {\it Kepler} \citep{Borucki2010}  have marked a milestone in stellar astrophysics, allowing the inference of stellar  properties (global parameters and internal structure) for a few thousands of stars, and opening the way to the synergy between asteroseismology and stellar population studies.
Due to different telescope characteristics and observation strategies,  the  asteroseismic databases  differ in spatial and frequency resolution, signal-to-noise level, and sampled regions of the sky. While {\it Kepler} observed during four years a 116 deg$^2$ field towards Cygnus-Lyra,  CoRoT collected 3-6 months light-curves  from more than ten 4 deg$^2$ fields in the Galactic disc in the Center and Anticenter directions.  K2 \citep{Howell2014}  has observed 19 fields close to the Ecliptic, during 75 days each.

The {\it Kepler} and CoRoT databases comprise currently more than 20000 G-K red giants (and 600 dwarfs and subgiants) for which at least the seismic global parameters \deltanu\ and \numax\  (and hence stellar mass and radius) have been measured \citep[e.g.][]{Yu2018, Mosser2010, Peralta2018, Chaplin2011a, Davies2016, Michel2008, Anders2017}. For a third of them,   period spacing ($\Delta P$) and coupling factor values have been also estimated  \citep[e.g.][]{Vrard2016, Mosser2017}. Finally, the internal rotation profile has been derived  for a few main-sequence pulsators \citep[i.e.][]{ Benomar2015, Aerts2015} and larger samples of evolved stars  \citep{Beck2012, Deheuvels2012, Deheuvels2014, DiMauro2016, Huber2013, Mosser2012}.  CoRoT data (and  K2), although  having lower frequency resolution and signal-to-noise than {\it Kepler} data, are good  enough to derive  \deltanu\ and \numax\,  as well as $\Delta P$ for some of the evolved stars \citep{Mosser2011}.

The quality of seismic data already collected sufficed to demonstrate their capability in inferring global and internal properties of stars, and in revealing hidden problems in current stellar modelling.  However, to unambiguously answer some of the outstanding questions in stellar physics we need to disentangle the contribution from  different quantities and uncertainties. A study of  simpler populations,  covering  different stellar parameters would be necessary.

{\it Kepler} observations of NGC~6791 and NGC~6819 -- two Galactic open clusters with ages of $\sim8$~Gyr and $\sim2.4$~Gyr, respectively -- have shown an enormous potential  clearly  revealing, for instance, the presence of non-standard evolution processes \citep{Handberg2017}. However, there are only four stellar clusters in the {\it Kepler} database, with the two young and sparse clusters NGC~6811 and NGC~6866 providing less stringent constraints than NGC~6791 and NGC~6819. In addition, the clusters M67 and M4 have been targeted for 70 days with K2, but such data do not allow detailed asteroseismic studies of stellar structures  
\citep{Stello2016, Miglio2016}. Overall, we have {\it detailed} asteroseismology for just two clusters, covering a small region of the age--metallicity plane. 

Moreover, the distribution of {\it Kepler} and CoRoT targets peaks at slightly sub-solar metallicity, and their selection functions are complex since they have been shaped by planetary science.
Unfortunately, the subsequent mission, NASA's TESS, although having the advantage of looking at much wider sections of the sky, has a much larger pixel size and shorter observation periods. 
It does not allow much more than global stellar properties to be obtained for the bulk of their targets \citep[see e.g.][]{Mosser2019, DaviesMiglio2016}. The upcoming ESA PLATO mission, instead, will have much improved signal-to-noise and extended observations for its two continuing view zones, allowing measurements of  individual frequencies and global parameters for more than 260000  targets \citep{ESA-SCI2017}. However, given the target selection aims mainly at finding nearby planetary systems characterizable through follow-up spectroscopy, PLATO detailed asteroseismic information will be available mainly for bright, nearby field F5-K7 dwarfs and subgiants. Moreover, the large pixel sizes of PLATO \citep[projected size in sky: 15 arcsec,][]{Samadi2019} will hamper obtaining detailed information for stars in crowded fields, including star clusters, the Milky Way's bulge, and nearby galaxies.

\section{Scientific objectives}
\label{sec:obj}
Having in mind the inferences on stellar structure possible with constraints from solar-like oscillations, and taking into account the limitations of present/planned asteroseismic missions, we identify a series of open, key science questions that a mission dedicated to high-precision, high-cadence, long photometric time series to dense stellar fields will be able to address.

\subsection{Science Objective 1: Unanswered questions in \textit{stellar astrophysics}}
\label{sec:SO1}
Research in stellar astrophysics is now reflourishing, thanks to the availability of the long-sought combination of high-precision spectroscopic and astrometric data, and to the advent of exquisite  asteroseismic constraints from space-borne telescopes.  However, while {\it Kepler} data have started exposing the limitations of current models, robust and more stringent tests of stellar structure are currently limited by the composite nature of the stellar population observed, and by the scarce number of low-metallicity stars with asteroseismic constraints.

High-precision tests of stellar physics need to be performed in ``controlled environments'', and clusters are the best stellar physics laboratories we have access to. 
Constraints from star counts, photospheric chemical abundances, and asteroseismology, together with strong priors on population-level parameters (e.g.  distance, initial chemical composition, age) will enable percent-level tests of predictions by stellar evolution, as presented below.  

Precise asteroseismic data for a set of open and, crucially, globular clusters will aid the theoretical development of a new generation of stellar models, which will benefit - in the next few decades - from an improved description of micro physics and 2D/3D numerical simulations of transport processes.  Such developments will have direct impact on most areas of astrophysical research, from the interpretation of the photometry of resolved galaxies \citep[see][]{Tolstoy2009} and integrated spectra of high-redshift galaxies \citep[see][and references therein]{Conroy2013}, to the detailed characterisation of exoplanets and their environment \citep[e.g. see][]{JCD2018}. 

Below we list open questions in stellar evolution which the proposed mission has the potential to address.

\subsubsection{Transport of chemical elements in the stellar interior} 
\label{sec:mixlowmet}

As described in detail e.g. in the recent review by \citet{Salaris2016}, the predictive power of current stellar models is limited by the simplified treatment of (and interaction between) various processes that concur at determining how chemical elements are transported throughout the stellar interior. 

Current shortcomings in the modelling of convection, atomic diffusion, mixing induced by thermohaline instabilities, gravity waves, and rotation affect the predicted evolutionary timescales, global stellar properties (e.g. surface temperatures, luminosities), internal structures, and surface abundances. These systematic uncertainties necessarily bias any model-based inference we make on e.g. stellar ages, also when seismic constraints are available. 

Moreover, the use of stellar photospheric abundances in studies of galactic chemical evolution relies on our understanding of internal mixing, as surface abundances may have changed compared to the initial value due to nuclear processes, and brought to the surface by internal mixing. Our limited understanding of such processes also affects the robustness of spectroscopic proxies of stellar ages like [C/N] \citep{Masseron2015,Martig2016, Mackereth2017, Casali2019},  which is based on how the efficiency of transport processes in convective and radiative regions depend on the mass and metallicity (hence age) of red giant stars  \citep{Salaris2017,lagarde2019}.

\paragraph{The properties of the convective boundary mixing layer}
All stars with masses above approximately 1.1 M$_\odot$ have convective cores during their MS, on the top of which a ``convective boundary mixing layer'' may develop. In classical 1D models of stellar structure, this layer is commonly known as the ``overshooting region''. Importantly, the efficiency of convective mixing (or overshooting extent in 1D models) affects the relation between the initial mass and MS age for all stars more massive than 1.1 M$_\odot$, or, alternatively, it affects all age determinations of stars younger than about 8 Gyr.
 
The best hopes to model the extent and structure of the convective mixing layer reside on modern 2D and 3D hydrodynamical models of stellar structure \citep[e.g.][]{Arnett2015, Baraffe2017}, which require robust observational constraints from asteroseismology. While, e.g. in red giants, there are a few approaches to measure the efficiency of convective mixing with {\it Kepler} data alone \citep{Montalban2013, bossini17}, they are complicated by a few factors, including e.g. the wide range of metallicities spanned by {\it Kepler} field stars, and its complex target selection. Detailed modelling of individual MS stars \citep[e.g.][]{Deheuvels2016, Michielsen2019} still has not produced unequivocal results. On the other hand, the distribution of gravity-modes period spacings in the {\it Kepler} clusters NGC~6819 and NGC~6791 seems to constrain quite well both the extent and the temperature gradient in the convective mixing layer of core-He burning stars \citep{bossini17}. The problem in this case is that present measurements regard just two clusters, and just the He-burning phase. 

Indeed, {\it Kepler}-like data for open clusters could provide crucial information for a more definitive assessment of the impact of convective mixing layer on stellar evolution (and hence the ``stellar clock'') as a function of stellar mass. Other advantages of extending the cluster sample, probably still not evident from present {\it Kepler} data, are: 1) Fast rotation is likely to  be the main contributor to the increased cores observed in post-MS phases \citep{Costa2019}; fortunately, this effect can be assessed via the observation of rotational velocities in the star clusters themselves, via either accurate photometry or high-resolution spectroscopy of stars on the MS \citep[e.g.][]{Cordoni2018, Bastian2018}. 2) In star clusters, the consequences of having larger/smaller cores can be directly observed also in the most advanced evolutionary phases -- including for instance its possible impact on the luminosity of early-AGB stars \citep{bossini15}, and on white dwarf (WD) masses \citep[i.e. the initial-to-final mass relation;][]{Cummings2019}. These can provide precious consistency checks for the overall efficiency of convective mixing and rotation, a step change compared to the loose constraints usually obtained from field stars.

\paragraph{Transport processes in radiative regions}
There is clear observational evidence for the occurrence of transport processes of chemical elements beyond convective instabilities. Several physical mechanisms have been proposed in the literature to explain such observations, including  atomic diffusion and radiative levitation, rotation-induced mixing, mixing by internal gravity waves, thermohaline mixing, and the complex interaction between these processes \citep[e.g.][]{Maeder2013, Mathis2009}.

For instance, helioseismology provides strong evidence that including atomic diffusion in the models reduces the tension with the seismically inferred sound speed profile  \citep[e.g. see the review by][]{JCD2002}. Evidence for diffusion from spectroscopy in clusters has also been reported -- e.g. in NGC~6397 \citep{Korn2007, Husser2016}, and from the APOGEE \citep{Souto2018} and Gaia-ESO \citep[e.g][]{smiljanic16,lagarde2019} surveys.
Albeit this evidence, a quantitative understanding and calibration of the efficiency of such processes, is still lacking.
Promising results on testing such effects on a star-by-star basis have been reported using high-quality asteroseismic data on main sequence stars \citep{Deal2017, Verma2019} and by combining seismic constraints on masses and spectroscopic observations in two open clusters \citep{Lagarde2015, Szigeti2018}.

The way forward is to explore these effects in stars in clusters, in particular in those where we expect to see an effect of diffusion changing with position on the  colour-magnitude diagram (CMD) such as, e.g., in Globular Clusters (GCs) and M67. Precise and accurate constraints on masses can be used to break the degeneracy between models with and without diffusion that affects the CMD \citep{Salaris2016}. Even more stringent constraints on the effect of diffusion on e.g. local chemical composition gradients, can be inferred for stars near and at the turnoff, providing an unprecedented calibration of the efficiency of transport processes. This is of paramount relevance if one wishes to improve the accuracy of age determinations for globular clusters (e.g. see \citealt{Chaboyer1992}), which are reference ``time pieces'' for the study of the early Universe (see also Sec.~\ref{sec:GCages}).

As mentioned earlier, one of the challenges that one has to face is to design observational tests that are specific to a given process. We know that for the case of rotation-induced transport processes, for instance, one would have to explain not only chemical composition profiles, but also the redistribution of angular momentum in stars. The latter can be probed by inferences on the rotational profile made using seismic constraints on large-scale velocity fields in the stellar interior, as illustrated below.

\subsubsection{Core rotation and transport of angular momentum}
\label{sec:rot}
The observations of solar-like oscillations in the Sun, in sun-like stars, and evolved stars have brought important constraints on their internal rotation rates \citep{Thompson1996, Benomar2015, Beck2012, Deheuvels2012, Mosser2012}. Comparisons between seismically inferred rotational profiles and models have demonstrated that an efficient transport mechanism is needed in addition to the transport of angular momentum (AM) by meridional circulation and shear instability \citep[e.g., see][]{Eggenberger2012,  Ceillier2012, Marques2013, Cantiello2014, Aerts2018}. 

Candidates for such an additional transport process include internal gravity waves and magnetic instabilities and their complex interplay with thermal and chemical gradient in the evolving stellar interior  \citep{Eggenberger2005, Cantiello16, Pincon2017, Eggenberger2019, Fuller2019}. First-principles, numerical simulations of such processes performed  under realistic stellar conditions are, however, particularly difficult to achieve, hence direct observational constraints on the internal rotation of stars are of prime importance to inform and test the modelling of angular momentum transport in stellar interiors. An improved understanding of AM transport processes is also key to understand the core rotation rates of compact objects \citep{Kawaler2015, Heger2005}.

To make further progress one needs more stringent and robust constraints on the evolution of AM transport, and to extend the constraints to the low-metallicity regime: this can be achieved by measuring rotational splittings of pulsation frequencies in stars in clusters. For instance, with accurate asteroseismic data e.g. on M67, we will have available direct observational evidence of how angular momentum is being redistributed along the evolution, from solar analogues through the MS turnoff, the subgiant phase and to the giant branches (see Sec. \ref{sec:targets}). 
Moreover, the detection of rotational splittings of mixed modes in stars in GCs will enable exploring metallicity effects on the transport of angular momentum, and testing models of AM transport in a regime which is closer to the one relevant for the first stars.

\subsubsection{Mass loss on the red-giant branch} 
\label{sec:massloss} 
While mass loss close to the tip of the RGB has been suspected for long, direct measurements of the mass loss efficiency are characterised by uncomfortably large uncertainties. The best determination so far seems to be based on asteroseismology, more specifically on the measurement of the mass difference between RGB and  red-clump (RC) stars in the clusters NGC~6819 and NGC~6791, observed by {\it Kepler}, and M67 using K2 data \citep{miglio12, Handberg2017, Stello2016}. These measurements however do not provide the dependencies of mass loss with key stellar parameters such as luminosity and metallicity, and do not even allow us to identify the basic mechanism responsible for the mass loss. Estimates of the mass loss in GC stars are also controversial, as demonstrated by the case of 47~Tuc: while its  horizontal-branch (HB) morphology requires stars to have lost about 0.2 M$_\odot$ on the RGB \citep[e.g.][]{Salaris16}, measurements both from spectra and infrared photometry do not give clearcut confirmation/disproval of mass loss \citep{Origlia2010, Boyer2010}, and have even been challenged by dynamical arguments \citep{Parada2016}.

On top of measuring the integrated RGB mass loss, oscillation frequencies can provide direct constraints on the mass of stars in the upper part of the RGB, where it is expected that the mass loss rates are the highest. The subclass of long-period variables known as OSARGs \citep{Soszynski2004} are potentially very useful in this regard.

OSARGs are compatible with stars in the RGB phase, just below the tip, and also with the AGB phase (\citealt{Tabur2010MNRAS}). Since their first detection it was suggested that they could be in fact solar-like oscillators \citep[see also][]{Dziembowski2010}. 
Recent works by \citet{Takayama2013} and \citet{Mosser2013} clearly support that OSARGs are the high-luminosity counterpart of solar-like oscillators observed by {\it Kepler} and CoRoT at lower luminosity. Although with smaller number of detected modes, the oscillation spectra of OSARGs are compatible with the presence of radial and non-radial acoustic modes, just as the low-luminosity RGB stars. The low radial order of the modes appearing in OSARGs and the low number of modes do not allow to define quantities such as \deltanu\ and \numax, however, 
as shown in \citet{Takayama2013}, combining the information on the sequence at which this dominant mode belong with the luminosity, would allow to characterise the mass of the star. 
Given their large amplitudes, and periods of few tens of days, these oscillations can be easily detected in all the clusters considered here.

In addition to mass loss, variations in the helium content $Y$ among the multiple stellar populations of GCs, are now believed to play a key role in determining the variety of observed HB morphologies. Therefore, a robust calibration of how much mass is lost on RGB stars at different metallicities, as allowed by asteroseismology, would contribute to open a clearer view into GC formation and evolution (Sect.~\ref{sec:helium}).

Robust measurements of integrated stellar mass loss, would also be key to determine accurate ages for the huge numbers of field core-He burning stars already observed by {\it Kepler} and K2, and those that will be observed by PLATO: while their present masses are provided by asteroseismology, the {\it initial masses} are needed for age determinations \citep[e.g.][]{miglio13,Casagrande2016, Anders2017}. Better estimates of mass loss on the RGB would also impact the initial conditions assumed for evolutionary models of AGB and post-AGB stars, used to characterise extra-galactic stellar populations.

\subsubsection{Occurrence of mergers / products of binary evolution} 
Thanks to constraints on stellar mass from asteroseismology in the open clusters observed by {\it Kepler}/K2, there is now clear evidence for over- and under-massive stars \citep{Brogaard2015, Brogaard2018, Stello2016, Handberg2017} which are the likely products of mass exchange with a companion. There is also likely evidence of such products among field stars that are enriched in $\alpha$ elements, hence supposedly belonging to an old population \citep{Martig2015, Chiappini2015, Jofre2016, Yong2016, Izzard2018, Hekker2019}.

The direct measurement of stellar masses could provide a better understanding of the observational abundance of blue stragglers and other types of rejuvenated stars, i.e. stars that previously accreted material from a companion star through a stable mass transfer event or a merger. An extension of the asteroseismic sample to a wide range of cluster ages and metallicities would provide invaluable constraints to the study of tidal interactions \citep[e.g][]{Beck2018}, and to 
 models of binary populations,  e.g. event rates of binary interaction, initial-mass-ratio/period distributions \citep[e.g., see][]{Moe2017} and constraints on key parameters in the modelling of interacting binaries.

For instance, robust inferences on the number of peculiar-mass stars will enable constraints on the $\alpha$ parameter that determines the efficiency of the transfer of orbital energy of the inspiraling cores to the envelope during the common-envelope phase of evolution  \citep[e.g., see][]{Izzard2018}. This parameter is the key to many processes in stellar astrophysics e.g. BHBH/BHNS/NSNS/WDWD mergers (hence sources of gravitational waves), type Ia supernovae, cataclysmic variables, FK Comae stars, X-ray binaries, post-(A)GB stars, and helium white dwarfs.

Moreover, testing the dependency with metallicity is also crucial. There are indications that binary interactions are more common at low metallicities \citep[e.g., see][]{Badenes2018}. Do we see that? Are the structures of the rejuvenated stars different at different metallicities? 

On top of characterising evolved stars that are products of mass transfer, blue stragglers near the  turnoff of old-open clusters will also be accessible to a precise asteroseismic inference, and will contribute to understanding e.g. how angular momentum is redistributed, so that it prevents the  accreting stars to quickly spin up to critical rotation (see also Sec. \ref{sec:rot}).

\subsection{Science Objective 2: Unanswered questions in \textit{cluster formation and evolution}}
\label{sec:clusterform}

\subsubsection{Globular clusters formation from absolute ages}
\label{sec:GCages}
Most GCs form in the first few Gyrs of the evolution of their host galaxy and they therefore contain valuable information about the earliest phases of galaxy formation. Because the stars in a GC are believed to have the same age, precise stellar (core) masses allow absolute age dating of GCs, with an age resolution that is sufficient to resolve the age differences among GCs.  
Combined with their metallicities, absolute ages allow us to distinguish accreted from in-situ formed GCs  \citep{forbes2010}, derive the accretion history of the Milky Way \citep{leaman2013,massari2019} and discover individual ancient ($\gtrsim10\,$Gyr) accretion events \cite[e.g.][]{myeong2018,myeong2019}. 
Accurate ages also serve as a valuable benchmark for other age dating methods, such as isochrone fitting and age dating from the white dwarf cooling sequence.

\subsubsection{Thresholds for the onset of multiple populations} 
Another unexpected trend is that only clusters older than 2 Gyr and massive enough, with a threshold close to 10$^5$ M$_\odot$ (e.g. \citealt{martocchia2019}), appear to host multiple populations (i.e. with the chemical signature) in the Magellanic Clouds. In the Milky Way only the really old and massive clusters, i.e. those classically called GC, show the presence of multiple populations (see \citealt{bragaglia2017} for a detailed list). Apparently, no open cluster, even among the oldest (e.g. NGC~6791, Berkeley~39), was massive enough at birth and/or formed in an environment favourable to the development of multiple populations. While these trends need to be confirmed by larger samples, the availability of asteroseismic masses/rotation for a few clusters will provide new light on this subject. 

\subsubsection{Measuring helium content in GCs with asteroseismology}
\label{sec:helium}

At present, the surface helium content $Y$ can be determined for giants in a GC from photometry using a combination of filters comprising the UV bands (essentially, only with HST, see e.g. \citealt{milone18}), or from the luminosity of the RGB bump \citep[e.g.][]{bragaglia10,lagioia18}. For HB stars, it can be derived modelling the extension of the HB, or directly though spectroscopy in a limited temperature range \citep[e.g.][]{villanova12,marino14}. In red giants, the only accessible He line is chromospheric, and its measurement is complicated \citep[e.g.][]{pasquini11,dupree13}. A further method to determine $Y$ for massive samples of giant stars in GC would then be very important.

As already mentioned in Sect.~\ref{sec:massloss}, a range of initial He abundances is mandatory to reproduce the HB morphology of old and intermediate-age stellar clusters and, in turn, to constrain the efficiency of mass-loss on the RGB \citep[e.g.,][]{Salaris06, Salaris16, Chantereau19, Tailo19}. 
To determine the mass of stars (through asteroseismology) is another indirect way to measure the He content since there is an anti-correlation between the initial He abundance and the initial mass of stars \citep[e.g.,][]{Salaris06,Chantereau15}. In addition, it is imperative to know the precise initial He abundance variation (and distribution) to greatly constrain scenarios trying to explain the origin of multiple populations. However, present determinations of the He content are still subject to large uncertainties and vary as a function of the method used. For instance, the determined He spread between \cite{Salaris16} and \cite{lagioia18} in 47~Tuc differ by a factor 3, making any progress in the multiple populations field difficult.

Therefore, {\it direct} asteroseismic constraints to $Y$ in cluster stars would be invaluable. In open clusters of different metallicity, they would also provide an empirical recipe for the He-enrichment law in the Galaxy.

\subsubsection{Globular clusters masses and dynamics} 
Precise masses of giants and an assessment of (photometric) binarity are valuable input for dynamical models of GCs. GCs are collisional systems and two-body relaxation  evolves the system towards  equipartition and  `mass segregation', such that the more massive stars and stellar remnants are more centrally concentrated.  
Dynamical mass models of GCs that include an equipartition description \citep[e.g.][]{dacosta1976, gunn1979, gieles2015} can be used to infer the underlying GC mass profile from the number density profile. For this, precise stellar masses help to `anchor' the models with multiple mass components. 
In addition, mass estimates at the different evolutionary  stages can be combined with dynamical models to estimate at which moments  the mass is lost \citep{Heyl2015,Parada2016}.
The properties of binaries are the result of the primordial binary properties and the dynamical evolution. For example, the inverse trend of the binary fraction with GC mass \citep[e.g.][]{milone2012} is likely the result of binary destruction in massive GCs. Photometric binary properties, complemented by the wealth of multi-epoch radial velocity surveys with Integral Field Units (IFUs) \citep[e.g.][]{kamann2018}, can be used to derive a full  census of binaries in GCs and can be included in dynamical modelling exercises of GCs. Finally, it allows us to understand the  interplay between binaries and the dynamical evolution of GCs and leads indirectly to insight in the formation properties (e.g. density and mass) of GCs.

\subsubsection{Black holes}
Based on an extrapolation of the `$M-\sigma$ relation' \citep{gebhardt2000,ferrarese2000}, GCs may host the sought-after intermediate-mass black holes (IMBHs, $\sim10^{3-5}\,M_\odot$), filling the gap between stellar-mass BHs and supermassive BHs. The existence of  IMBHs is highly debated, and from radio and X-ray observations, stringent upper limits of $\sim10^3~M_\odot$ were found for several Milky Way GCs \citep{tremou2018}. However, GCs are gas free and the absence of an accretion signal may not yet be evidence for the absence of IMBHs.
Various claims have been made for an IMBH in $\omega$~Cen \citep[e.g][]{noyola2008, baumgardt2017}, 47~Tuc \citep{kiziltan2017},  NGC~6397 \citep{kamann2016} and M54 \citep{Ibata2009}. These findings are based on kinematics of stars in the centre of the cluster, and dynamical mass modelling. A complication with this method is that the IMBH signal is (partially) degenerate with other -- more plausible -- effects,  such as radial orbit anisotropy \citep[e.g.][]{zocchi2017} and a population of stellar-mass BHs \citep{zocchi2019,lutzgendorf2013}. An alternative approach is to look for gravitational-lensing effects \citep{kains2019}. In the search for IMBHs, various stellar-mass BHs have been reported in several GCs \citep[e.g.][]{strader2012}, including 47~Tuc and $\omega$~Cen. High-precision and high-cadence photometry could be used for micro-lensing events and light-curve modelling of BHs with  stellar companions \citep[see][for an example in NGC~3201]{giesers2018} to more firmly establish the presence of stellar-mass BHs in GCs, which impacts on our understanding of the role of GCs as gravitational wave factories. 

\subsubsection{Do stars in clusters rotate faster than field stars?} 
One of the most surprising results of high-precision HST photometry of Magellanic Cloud clusters regards the presence of a large fraction of main sequence stars rotating close to their critical break-up velocities, $\Omega_\mathrm{c}$. For instance, in the $\sim200$ Myr LMC cluster NGC~1866, about 2/3 of the MS stars appear to be rotating at $\sim0.9\,\Omega_\mathrm{c}$ \citep{milone17}.
Moreover, extended main sequence turn-offs (MSTOs)  are an ordinary feature in intermediate-age star clusters \citep{Milone2009, Cordoni2018} and probably largely due to the presence of fast rotators \citep{bastiandemink, Bastian2018}. 
Although the distributions of rotational velocities in field stars \citep{Royer2007} is more difficult to interpret, projected rotational velocities seem to indicate that young clusters have more fast rotators than their neighbouring fields \citep[see][]{Huang06, Strom05, Dufton06}. 
Therefore, the question arises whether this is a general trend, applying to all stars born in clusters. Some authors \citep[][see also Sec. \ref{sec:spin}]{Kamann2019, corsaro17} also suggest that cluster stars largely inherit the initial angular momentum of the molecular cloud from which they form, hence spinning faster and at preferential orientations as compared to the field. These suggestions could be verified with more data on rotational velocities added to the asteroseismic information (rotational velocities and spin orientations), in and around a few star clusters. %and in their neighbouring fields.

Given that stellar evolution models are usually calibrated with observations of star clusters, and later applied to all galaxy components irrespective of their density, this question may be very important in the context of stellar evolution and stellar population synthesis of galaxies. A very basic assumption of these models, namely that stars in clusters form and evolve in the same way as stars in the field, may be wrong.

Faster rotation can also arise from more mergers among young stars in a dense cluster environment, which may lead to more rotating (proto)stars. This  might be worth investigating with the additional constraints (including stellar masses and core rotation rates) provided by asteroseismology. 

\subsubsection{Spin alignment in clusters} 
\label{sec:spin}
Recent results have suggested that the alignment of stellar rotation axes in clusters is not isotropically distributed, but rather has a preferential inclination angle \citep{corsaro17}. This result is based on asteroseismology analysis of {\it Kepler} data for stars in NGC~6791 and NGC~6819. %, two Galactic open clusters with ages of $\sim8$~Gyr and $\sim2.4$~Gyr, respectively.  
The generally advocated explanation for the alignment of stellar rotation axes in clusters is that the stars were born within a giant molecular cloud that contained significant amounts of angular momentum, which was inherited by the stars as they formed \citep[e.g.][]{corsaro17,Reyraposo18}.
While some caution needs to be taken in interpreting these results, due to potential systematic effects in inferring the stellar inclination angle from \textit{Kepler} analyses \citep{Kamiaka2018, mosser18, Kuszlewicz2019}, the results potentially offer an unexpected and intriguing insight into the formation of stellar clusters. Looking at these two clusters, \citet{Kamann2019} found that the cluster NGC~6791 rotated with an axis that was consistent with the spin alignment of the stars, suggesting that angular momentum has transferred from the large scale (cloud/cluster) to stars during the early phases of formation.  Expanding this type of study to a statistically meaningful sample would be a major step forward.

\subsection{Science Objective 3: Assembly history and chemodynamics of the {Milky Way} and dwarf galaxies}
\label{sec:SO3}

One of our main goals is to improve the quantity and quality of asteroseismic constraints at low metallicities, with new GC observations. This alone would produce an improvement in stellar models at low metallicity (see e.g. Sec.~\ref{sec:mixlowmet}), possibly impacting the interpretation of data for low-metallicity dwarf galaxies, and the resolved halos of more massive galaxies, presently observed out to the Virgo cluster. 
Even more certain would be the impact coming from the {\it direct} observation of a few galaxy components, such as:
  
\subsubsection{$\omega$~Cen} 
Rather than a normal globular cluster, $\omega$~Cen is probably the heart of a dwarf galaxy whose periphery has been dispersed by the Milky Way. It contains multiple stellar populations, with a large range of metallicities (i.e. heavy element content) that betray a formation over an extended period of time. 
Interestingly, $\omega$~Cen is close enough for asteroseismic observations to reveal accurate masses of its RGB and red-HB stars (with long time series allowing even the detection of internal rotation; see Fig.~\ref{fig:detect}). Given the available  HST multi-band photometry and spectroscopy \citep[e.g.][]{Hilker2004, Piotto2005, Villanova2014}, the accurate seismic masses would allow a much more complete picture of the star formation and chemical evolution history in this unique object \citep{Romano2007}.

\subsubsection{The Sagittarius Dwarf Spheroidal and M54} 
Among the many dwarf spheroidal galaxies, the Sgr dSph is both one of the closest, and more interesting ones. It contains a nuclear star cluster at its core, M54, which is expected to evolve similarly to $\omega$~Cen \citep{Carretta2010}. Its distance (25 kpc away) represents the main difficulty for asteroseismic observations; however, our simulations (Sec.~\ref{sec:seismic_perf}) indicate that solar-like oscillations are detectable at the level of the RC and above, provided a mirror with diameter $\gtrsim90$cm and at least 6 months of observations. These observations could represent the first asteroseismic measurements for an external galaxy. They would allow a determination of the Sgr dSph star formation history, which then may be used to infer possible age gradients and to date when Sgr dSph entered within the sphere of influence of our Galaxy.

\subsubsection{Magellanic Clouds}
The LMC and SMC galaxies have always been fundamental testbeds for the theories of stellar evolution and stellar populations. One could conceivably detect solar-like oscillations in the upper RGB, provided a large enough mirror size could be considered (see Sec.~\ref{sec:seismic_perf}). 
In the context of the proposed mission, one could run a pilot asteroseismic study of the Magellanic Clouds. For instance, this is easily done in the case of the SMC, which lays in the background of two populous GCs, namely 47~Tuc and NGC~362.
{This would, at the very least, enable to extend studies based on large-amplitude pulsators such as OSARGs (\citealt{Soszynski2007}, see also Sec. \ref{sec:massloss}) to stars with lower luminosities and lower pulsation amplitudes.}

\subsubsection{The {Milky Way} bulge}
\label{sec:bulge}

The {Milky Way} bulge still holds many unanswered questions despite the modern data \citep[see][for a recent review]{Barbuy2018}: not only is it a region where several galactic components co-exist (spheroidal, bar, innermost parts of the thick and thin disc, and residuals from merger accretions), but it is also not easily reachable given the large extinctions, large distances involved as well as crowding. Regarding its spatial structure and kinematics, added to the general idea of it being a box/peanut bulge, there are still many uncertainties regarding, for instance, the fraction of bulge which is actually in a bar, the different spatial distributions of metal-rich and metal-poor components, and the nature of its X-shaped feature \citep[see e.g.][]{Zoccali2017}. Regarding its stellar populations, its age distribution is still a matter of debate, with some observations suggesting that the {Milky Way} bulge is not simply ``old'' but contains  a significant fraction of intermediate-age populations \citep[e.g.][]{Saha2019}, whose fractions are far from being established. For instance, while the precise HST photometry from \citet{Brown2010} indicates only old stars, the spectroscopic masses of microlensing dwarfs from \citet{Bensby2017} reveal a $\sim$15\% fraction of young stars. 

While large spectroscopic surveys are providing excellent kinematics and chemical abundances, the weak point of present bulge studies resides in the lack of accurate distances and ages. For distances, even by combining Gaia DR2 and APOGEE DR16, the Bayesian spectrophotometric code StarHorse (\citealt{Anders2019}, \citealt{Queiroz2020}) still has distance uncertainties of the order of 10\% in that region -- 
therefore not only insufficient to disentangle the co-existing stellar populations in the bulge region, but also to get  large enough samples of stars.
The principal difficulties stem from the high extinction values, its unknown distribution along the line-of-sight, and the position-dependent non-standard extinction curves \citep[e.g][]{Nataf2016,Saha2019}.  
Such problems will persist even with Gaia parallaxes, which will not be able to peer through the dust to reveal the real stellar distances over most of the bulge. An even more incomplete situation is the one for age estimates in the bulge region. In this specific front, asteroseismology can be of great help, by providing accurate masses and radii of bulge red giants, and hence their ages and distances. 
Combined with Gaia's proper motion data, these data enable us to map the age of stars depending on the orbits. Stars in the bar structures, such as the long bar and the shot X-shaped feature, are in the trapped orbit, and the age distribution of stars in the orbit of each feature tells us the formation epoch of each component of the bar structure. Recently, \citet{Bovy2019} made the first attempt to age date the stars in the Galactic bar and claim that the bar is dominated by old stars, and therefore formed at the early epoch. However, this is relying on spectroscopic distances and ages, which are calibrated with the currently limited asteroseismic data, and rely on current stellar model tracks. Direct and precise seismically inferred ages and distances of the bulge stars will unambiguously  answer the formation epoch of the bar, even the difference of the formation epochs of the long bar and the X-shaped feature, which is considered to form by ``buckling'' that occurred more recently. 

Of course, asteroseismic observations of the bulge should also meet some basic requirements, such as being sensitive to redder wavelengths and avoiding too-crowded regions. A careful field selection will be needed in this case, and will be aided by combining Gaia data with ongoing/near-future spectroscopic surveys (such as 4MOST, APOGEE South, and MOONS). 
The observations would be complementary and have a clear synergy with the recently selected Japanese-led {\sc Small-JASMINE} mission (planned launch in mid 2020s, \url{http://jasmine.nao.ac.jp/index-en.html}), which will provide near-infrared astrometry and time-series photometry data in the Galactic nucleus region (within $\sim\!0.7^\circ$).

\subsection{Additional/complementary science from high-cadence, high-precision long photometric monitoring}

\subsubsection{Solar twins and ``solar evolution'' from M67 observations}
With its solar metallicity and $\sim$4~Gyr age, the star cluster M67 can provide extraordinary insights on the evolution of the Sun. It contains a handful of ``solar twins'', stars whose spectra are very close to the Sun's \citep{Pasquini2008} and which indicate a nearly-identical chemical composition \citep{Liu2016}. Asteroseismic analysis of these stars is granted to provide an interesting comparison with helioseismology. In addition, M67 could offer additional, population-level constraints on the evolution of rotation, activity, and  diffusion, in stars which were initially similar to the Sun. As discussed in Sec.~\ref{sec:targets}, M67 is close enough to allow detailed asteroseismic analysis of MS stars in addition to its numerous red giants. Given its position close to the Ecliptic, PLATO observations of M67 (which are still to be defined) will be limited to short time series, and to the cluster's outskirts. 

\subsubsection{Exoplanets in clusters / bulge}
Characterising planetary systems in star clusters is crucially important to constrain our theories on planetary formation and subsequent migration. Clusters provide us with a homogeneous environment where fundamental parameters such as chemical abundances and age can be reliably measured, in a much better way than how it is done on field stars, putting the detected planets in an accurate evolutionary context.
While there is a growing number of exoplanet surveys focusing on low mass stars and binary systems, only very few attempts were, and are being made to explore alternate Galactic environments. So far most exoplanetary systems have been detected orbiting single, Sun-like stars, with only a handful of planets detected in open clusters by {\it Kepler}, K2 and TESS \citep{Fujii2019}. Quite surprisingly, the occurrence of planets within open clusters (OCs) has been suggested to be similar or even slightly higher than for field stars, although the small sample size severely limits the statistical significance of the results \citep[][the latter found a $5.6_{-2.6}^{+5.4}$\% frequency of hot-Jupiter planets  around solar-metallicity stars]{Meibom2013, Brucalassi2016}. Unfortunately: 1) only four OCs were located within the {\it Kepler} field, all of them too distant for a planet search to be fully effective on main sequence stars; 2) K2 and TESS were/are not particularly helpful in enlarging the sample, given their short baseline ($\sim$ 70 and 27 days, respectively) that limits the discovery space mostly to hot planets; 3) PLATO will be limited by its large pixel scale ($\sim 15"$, only slightly smaller than TESS), which implies that confusion will badly impact its photometric performances on crowded fields, also boosting the occurrence of false-positive detections. Confirming (or disproving) a higher rate of hot Jupiters in OCs would favour (or reject) the \textit{planet-planet scattering} scenario, which is one of the competing theories to explain the origin of hot Jupiters \citep{Dawson2018}. On the other hand, evidence for small (rocky) planets in OCs is still fragmentary and to be explored.
      
The situation is even less clear when it comes to globular clusters. For instance, HST observations of 47~Tuc revealed no transiting hot Jupiter  \citep{Gilliland2000} and also subsequent efforts on $\omega$~Cen and NGC~6397 were unfruitful \citep{Weldrake2008, Nascimbeni2012}. The interpretation of these results is difficult. It remains unclear whether this absence is caused by the low metallicity of the cluster, by the intense density of stellar systems (which could either reduce disc and therefore planet formation, or could lead to system-wide instabilities post-formation), or both. It is also debated whether the null result on 47~Tuc is actually statistically significant \citep{Masuda2017}. 

Microlensing surveys sometimes detect exoplanetary systems into the bulge \citep{Zhu2017}, but those planets usually orbit fairly far from their host, preventing any direct statistical comparison to transit and Doppler surveys. Altogether, this calls for additional, more detailed investigations. In other words, we need a new mission to constrain the Initial Radius/Mass Function for exoplanets in different (and largely unexplored) environments.

A new {\it asteroseismic} mission, conducting photometric observations as precise as {\it Kepler} in the Milky Way's bulge, and clusters of various ages and metallicities, would beautifully fulfil this role, by being able to detect transiting planets within these multiple stellar environments. It would allow us to verify whether the statistics of planet detection match one another between each of these environments, and whether they match the detections of {\it Kepler}, {\it TESS} and of all the ground-based results. It would allow us to seek similarities and differences caused 1) by environment (high stellar densities might lead to more frequently unstable systems, decreasing how packed planetary systems are found \citep[e.g.][]{Malmberg2011}, 2) by metallicity (confirming how it correlates to the presence of gas-giants and of super-Earths \citep[e.g.][]{Sousa2019}, 3) by metal abundances (asking whether $\alpha$-enhanced stars can produce a different planet population), 4) by age (since we expect planet populations to change with time \citep{Pu2015}, and 5) by history (by tracking systems that have migrated within the Milky Way, moving from one environment to another). 
The length of each stare will determine up to which orbital period transiting planets can be detected. Microlensing might also be attempted. With enough transits, masses for exoplanets can be determined from transit-timing variations. Foreground white dwarfs, and very late M dwarfs might fall within each field, with the potential to discover Earth-sized planets. Particularly interesting would be the possibility of discovering ``Solar System twins'' orbiting the M67 solar twins. 

\subsubsection{Activity / flares}
 It has long been established that solar p-mode frequencies respond to the Sun’s changing levels of magnetic activity and hence provide a probe of the underlying physical changes driving these variations, i.e. stellar dynamos. Seismic detections of stellar-cycle-related variability are now being extended to other solar-like oscillators thanks to precise, long-timebase photometric data from CoRoT and {\it Kepler} (see Sec. \ref{sec:preamble}). However, these detections are in field main-sequence stars. Here, there is a great scientific opportunity in extending studies to counterparts in markedly different stellar environments.  The additional constraints provided by common cluster membership opens the exciting possibility to probe the action of stellar dynamos as a function of stellar mass.
 
Stellar activity is likely to have a large influence on the habitability of exoplanets. Therefore in order to understand the chances for finding life in the Universe, we need to understand the nature and evolution of activity of Sun-like stars. Over the last decades we have learned that cycles like the 11-year sunspot cycle are common in Sun-like stars \citep{1995ApJ...438..269B}, that young stars tend to have enhanced levels of UV and and X-ray emission \citep{2007LRSP....4....3G} and that some stars, including Sun-like stars, host devastating superflares \citep{2012Natur.485..478M}. 
 
The proposed mission will allow us to study the photometric effect of stellar activity in a large number of stars with well constrained parameters. This will be especially important for the studies of superflares as it is still an open question if the Sun would be capable of hosting superflares \citep{2016NatCo...711058K, 2019ApJ...876...58N}. In order to answer this question we need to either detect or rule out superflares on truly Sun-like stars with respect to mass, age, rotation, and chemical composition. Here focusing on stars in stellar clusters will be a great advance. 

Finally, the high-precision, high-cadence photometric monitoring of stars along the HB in GCs, will also shed light on the nature of the variability of blue horizontal branch (BHB) stars, and its relation to chromospheric activity / chemical peculiarities \citep[see e.g.][]{Paunzen2019,Montenegro2019}.

\subsubsection{Additional science from stellar pulsations}
The driving of pulsation modes occurs typically near the surface, and is therefore mainly a function of the effective temperature. The part of the HR diagram going from the classical instability strip to the red-giant branch (main targets of the proposed mission) corresponds to stars with a convective envelope. By one way or another, convective motions and ionization zones always drive pulsations. This is clearly seen on the main sequence: going from blue to red, we encounter $\delta$~Sct stars (HeII ionization zone driving), $\gamma$~Dor stars \citep[driving at the base of the convective envelope, see e.g.][]{Dupret2005} and solar-like oscillators (stochastic driving by convective motions). Going up to post-main sequence stars we observe on one side RR Lyrae and Cepheids and on the other side solar-like oscillations of red giants. There is no obvious reason for the absence of pulsation in-between. The only obstacle to non-radial pulsations in evolved stars is radiative damping in the core. However, partial reflections in the evanescent zone usually leads to the trapping of some non-radial modes in the envelope, making their detection possible. As long as the core-envelope density contrast is not too high, mixed modes probing deeper layers are also predicted \citep{Dupret2009} and observed \citep{Bedding2011} in red giants, increasing considerably their seismic potential: for instance, probing of the core rotation and density. Note also that the instability strips associated with different types of mode driving processes always intersect, producing hybrid stars with even higher seismic potential. We expect mixed modes and hybrid stars to be also present in hotter stars, making their very detailed seismic characterization possible for the first time.

\subsubsection{Additional targets in the field} 
\label{sec:othertargets} 
Asteroseismic surveys carried out in the 2020--2030 era will benefit from extensive new data from wide-area surveys such as Gaia, eROSITA, the Rubin Observatory (previously known as the LSST), SDSS-V, which will provide a wide variety of new targets for more detailed time-series, high-cadence photometry, both within and outside star clusters. It is hard to foresee all the possibilities here, but we might expect:
\begin{itemize}
    \item better characterization of classical pulsators, eclipsing binaries, and close binary systems uncovered by ``low-cadence surveys'' such as Gaia and the Rubin Observatory; 
    \item improved statistics of FK Com stars, which are rapidly rotating giants probably resulting from recent stellar mergers. {\it Kepler} allowed direct measurements of rotation periods for a handful of them \citep{Costa2015} and confirmed their association with bright X-ray sources with high levels of chromospheric activity and flaring behaviour \citep{Howell2016}, hence narrowly opening the way to link models (of individual objects or of populations) to their observations.     \item follow-up lightcurves of unusual transients, such as those found by {\it Kepler} and which still remain without satisfactory explanations \citep[e.g. KIC~8462852;][]{Boyajian2016}.
\end{itemize}

\section{Translating top-level science objectives into target selection and a possible mission profile.}
\label{sec:targets}
Given the main scientific objectives outlined in Sec.~\ref{sec:obj}, we discuss here the targets needed to achieve our goals and give indications about the key factors that will concur to define the mission profile.
\subsection{Targets}
 We make a preliminary selection of suitable targets considering proximity and diversity as the main selection criteria.
Table \ref{tab:targets} gives an example of a target list, which includes three globular clusters (with different chemical compositions, morphologies of the HB, and properties of multiple populations), $\omega\, \rm Cen$ (likely the nucleus of a satellite galaxy, with extended star formation history), an old open cluster (M67), and a field with a significant bulge population. We have also listed the Sgr dwarf spheroidal galaxy among the targets worthy of consideration.

\subsection{Seismic performance}
\label{sec:seismic_perf}
As presented in \citet{Mosser2019}, the plethora of data and analyses performed on photometric time series from several space-borne telescopes allows us to define and quantify a robust seismic performance indicator. For the targets in  Table \ref{tab:targets}, we adopt the method described in detail by \citet{Mosser2019} to quantify the performance expected when varying the mirror size and duration of the observations. 
The latter are the two key parameters impacting on the seismic yields, provided e.g. that the pixel size of the detectors on the sky is such as to mitigate the effect of contamination from nearby sources\footnote{Depending on the specific choice of the target, a pixel size of the order of 1" or less is needed, compared to 4" of {\it Kepler}, 20" of TESS and 15" of PLATO} and a cadence appropriate for the stars of interest is chosen. The noise model, as described in \citet{Mosser2019}, includes a systematic component originating from e.g. the sky background, the jitter, or the readout noise, and it is assumed to be similar to {\it Kepler}'s. We assume a duty cycle of 95\% and consider a broad, white-light photometric filter. 
%\begin{landscape}
\begin{table}[]
    \caption{Approximate characteristics of the targets for which we assess the seismic performance in Sec. \ref{sec:seismic_perf}. This list is indicative only, alternative options are available.}
    \label{tab:targets}
    \vspace{.25cm}
    \centering
{\rowcolors{2}{grigioc!10}{white!0}
\tabcolsep5.5pt
{\renewcommand{\arraystretch}{1.2}% for the vertical padding
\begin{tabular}{ l|c|c|c|c|c  }
\hline
%\multicolumn{3}{|c|}{Clusters} \\
%\hline
Object name & d [kpc] & $m_{\rm V,\, RC/HB}$ &[Fe/H] & Age [Gyr] & ang. size \\
\hline
NGC 104 (47 Tuc) & 4 & 14  &  -0.8 & 13 &  $31'$\\
NGC 6121 (M4)  & 2.2 & 13 &  -1.5 & 12.2 &   $26'$\\
NGC 6397 & 2.4 & BHB  &  -1.8 & 13.4  & $32'$\\
NGC 2682 (M67) & 0.9 &  10.5  & 0.0 & 4.5 & $30'$ \\
\hline
NGC 5139 ($\omega$ Cen) & 4.9 & 14 & broad   & broad  & $36'$ \\
Milky Way's bulge &  6--10  & $I>14$ & broad &  broad & wide \\
Sgr dSph & 25 &  18--20  & broad & broad & $450' \times 216'$ \\
%\hline
%LMC & 50 kpc &  &   &  &  & & &\\
%SMC & 60 kpc &  &   &  &  & & &\\
\hline
%\hline
%Andromeda - just for ref & 700 & 24? &   &  &  & & & \\
\end{tabular}
}
}
\end{table}

\begin{figure}
\centering
\begin{minipage}{.5\textwidth}
  \centering
  \includegraphics[width=\linewidth]{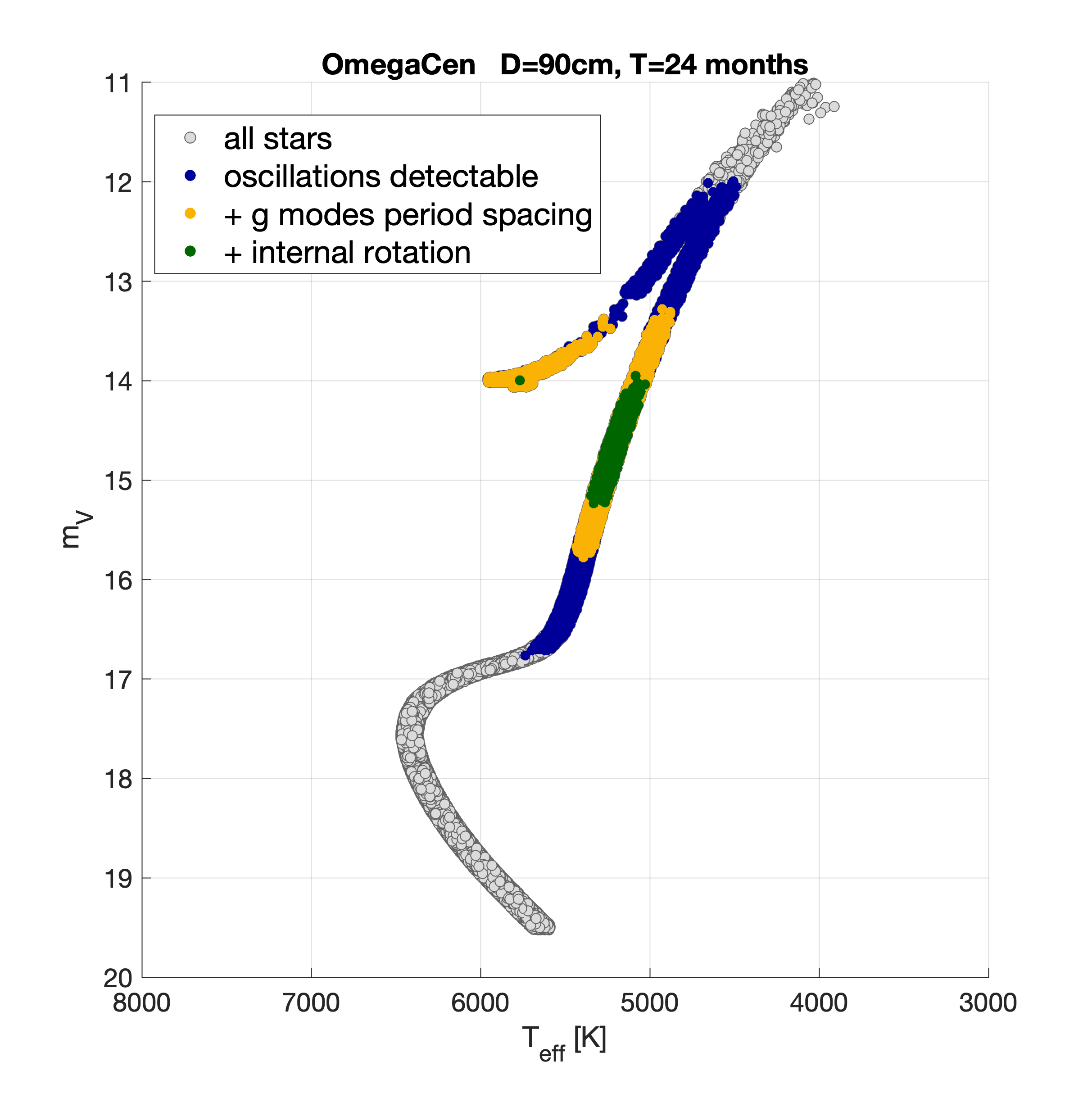}

\end{minipage}%
\begin{minipage}{.5\textwidth}
  \centering
  \includegraphics[width=\linewidth]{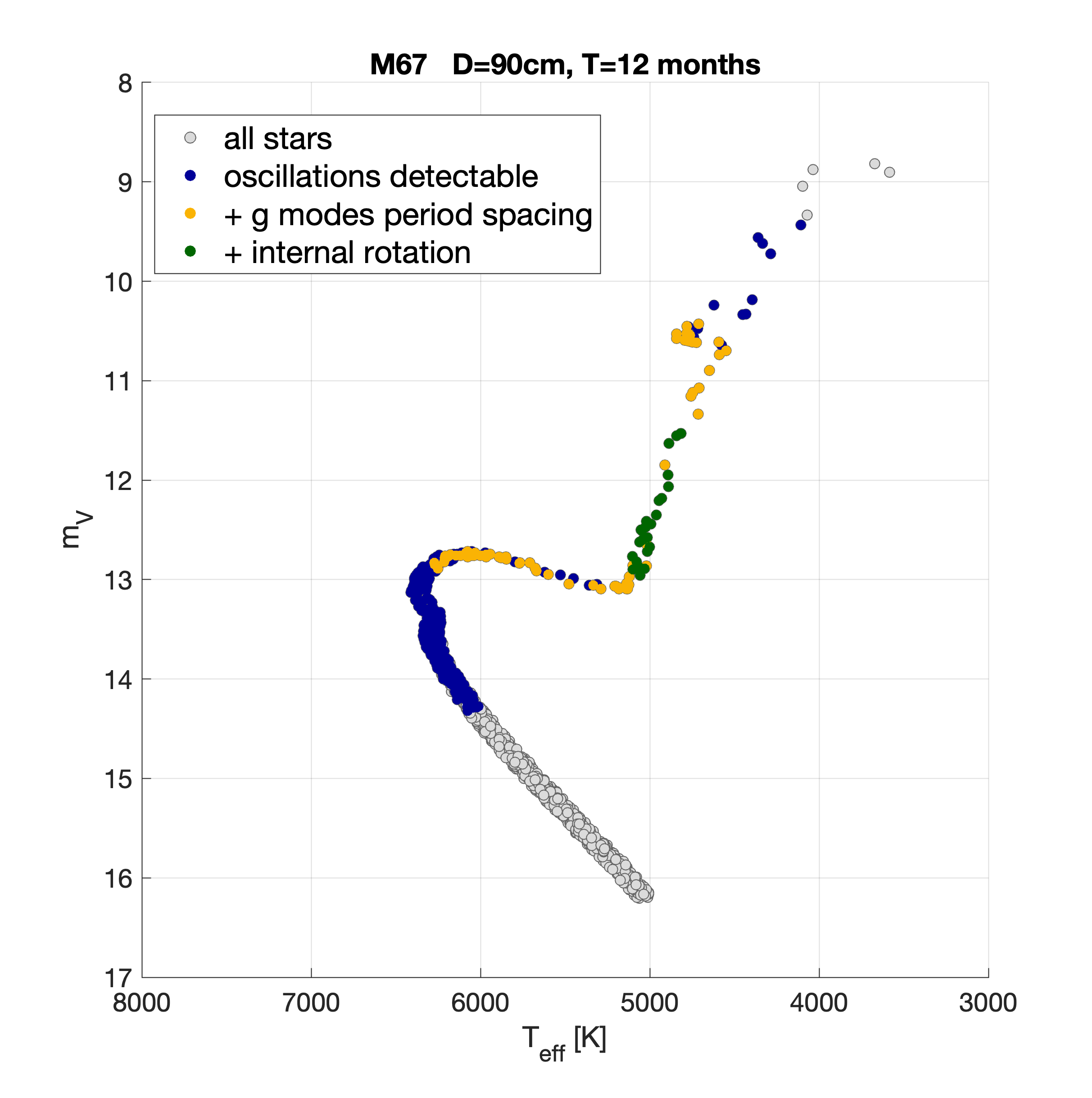}

\end{minipage}

\caption{Hertzsprung-Russell diagram showing stellar populations representative of $\omega$~Cen ({\it left panel}) and M67 ({\it right panel}). Each star in the population is coloured according to the seismic information that can be extracted: blue, solar-like oscillations are detectable; yellow, information on the gravity-mode period spacing can also be inferred; and green, rotationally split pulsation modes can be measured, hence information on the internal rotational profile can also be inferred. We assume a mirror diameter of 90cm and a duration of the observations of 2 years ($\omega$ Cen) and 1 year (M67). }
\label{fig:detect}
\end{figure}

\begin{figure}
    \centering
\begin{minipage}{.5\textwidth}
  \centering
  \includegraphics[width=\linewidth]{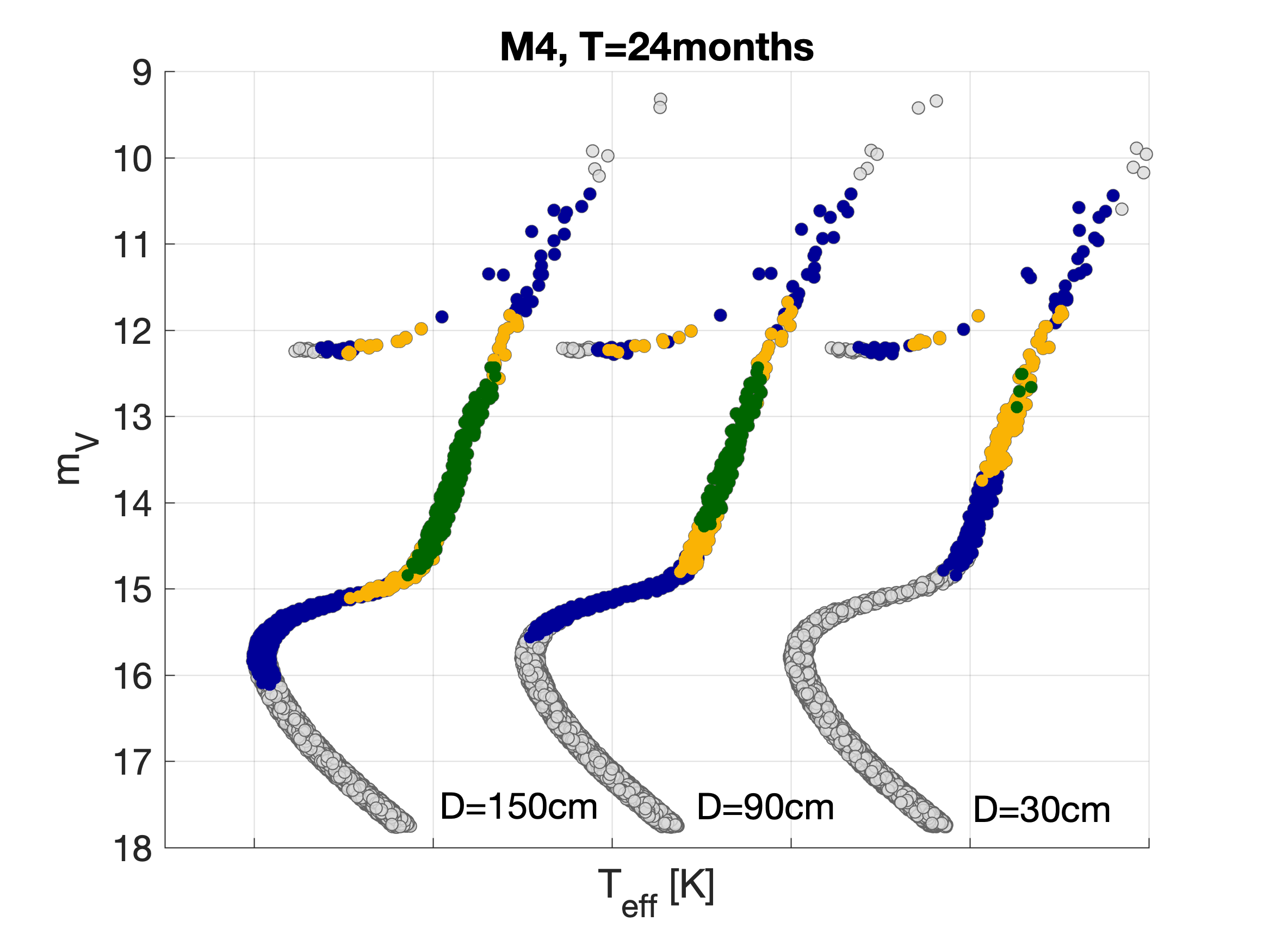}
\end{minipage}%
\begin{minipage}{.5\textwidth}
  \centering
  \includegraphics[width=\linewidth]{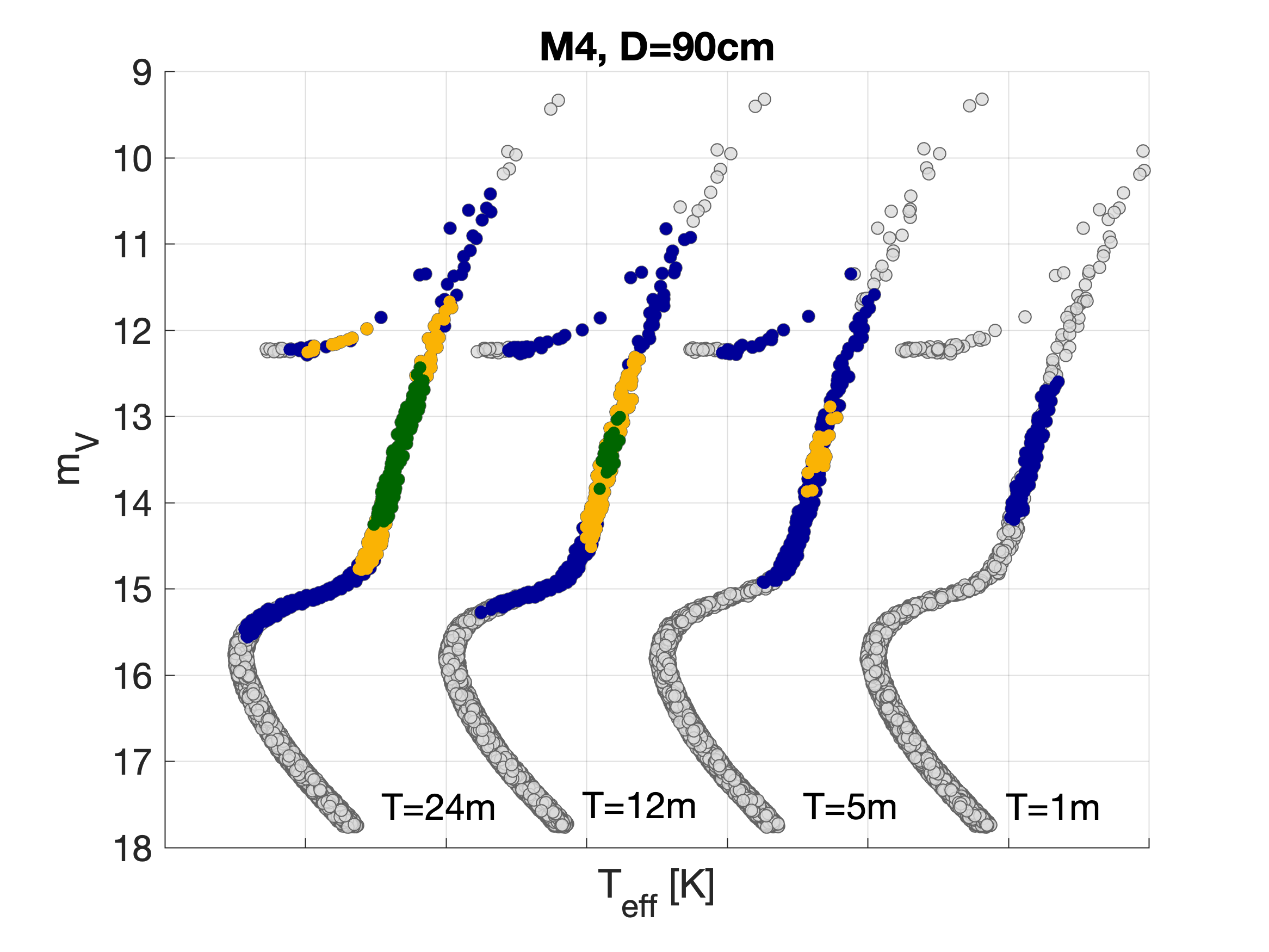}
\end{minipage}    
    \caption{Same as Fig. \ref{fig:detect}, this time for one cluster (M4) but changing the diameter of the telescope ({\it left panel}) or the duration of the observations ({\it right panel}).}
    \label{fig:detectDT}
\end{figure}

For the targets listed in Table \ref{tab:targets} we quantify our ability not only to detect solar-like oscillations, but also to measure the detailed properties of mixed modes, and to resolve rotationally split frequencies, which are needed e.g. to achieve the science objectives presented in Sec. \ref{sec:SO1}. To simulate the stellar population in the proposed fields, we use {\sc TRILEGAL} %/CMD}%\footnote{\url{http://stev.oapd.inaf.it/cgi-bin/cmd}}
\citep{Girardi2005} with literature-based estimates for the age, chemical composition, distance, and extinction for each target. Since the seismic performance indicator is largely  determined by global stellar properties and the targets' distance, the dependence on the details of the stellar models used to simulate the populations is not a source of concern.

Examples of the detection thresholds obtained with a different combination of telescope size and duration of the observations, for some of the key targets in Table \ref{tab:targets}, are illustrated in Figs. \ref{fig:detect} and \ref{fig:detectDT}. 
We present three cases: a telescope with a diameter of 30cm, 90cm, and 150cm and vary the duration of the observations from 1 month to 2 years. Since our aim is not only to infer global stellar properties (e.g. mass) but, at least for some of the targets, to perform high-precision tests of stellar structure models, we argue that a collecting surface corresponding to a $\sim 90$cm diameter and a combination of 24-, 12-, and 6-months long campaigns for the candidate targets in Table~\ref{tab:targets} would enable reaching the key objectives presented in Sec. \ref{sec:obj}. 
For instance, if one aims at a precise and accurate characterisation of the rotational profile and to infer detailed, local properties of stellar interiors using the information from individual mixed p-g modes, observations longer than $\sim1$ yr are needed (e.g. Scientific Objective 1, Sec. \ref{sec:SO1}). On the other hand, 6-months long observations are enough in most cases to enable robust and precise inferences on global stellar properties (e.g. masses, ages) to a level which would enable breakthroughs in studies of stellar populations (Scientific Objective 3, Sec. \ref{sec:SO3}). 

Larger mirrors would be needed if one wishes to fully characterise giants at and below the luminosity of the RC in satellite galaxies, e.g. in the LMC ($\mu\simeq18.5$). 
A small-size mirror (e.g 30cm) would still enable detection and characterisation of stars in the closest globular clusters, yet limit the high-precision tests of stellar physics, especially at the turnoff of globular clusters, and on the MS of open clusters.

\subsection{Preliminary considerations on telescope design}
As evinced from the typical angular size of our prime targets (see  Table \ref{tab:targets}), an appropriate diameter of the field of view is $\lesssim 1 \deg$. A suitable field for the bulge will be chosen based on detailed studies enabled by Gaia and in synergy with 
{\sc Small-JASMINE} (Sec.~\ref{sec:bulge}). 
Similarly, for the Sgr dSphe, a suitable sub-field can be identified by a detailed study of its stellar content and foreground contamination.

As discussed in Sec.~\ref{sec:seismic_perf}, a 90cm aperture telescope would allow us to reach our main goals in terms of seismic performance. 
Assuming pixel sizes of about 5$\mu$m, the telescope focal length will depend on the adopted pixel scale. Pixel scales of the order of 0.25 arcsec/px (to be compared to the 4 arcsec/px of {\it Kepler}) are certainly well suited to separate stars in open clusters and in the outskirts of globular clusters and the bulge. They would imply a focal ratio of about 5, hence a reasonably-compact telescope design. However, detailed simulations (using available images of our targets) are necessary to assess the optimal pixel scale for all targets, considering all the other constraints on telescope dimensions and weight.

\section{Summary}
In the last decade, the \textit{Kepler} and CoRoT space-photometry missions have demonstrated the potential of asteroseismology as a novel, versatile, and powerful tool to perform exquisite tests of stellar physics, and to enable precise and accurate characterisations of stellar properties, with impact on both exoplanetary and galactic astrophysics. 

Based on our improved understanding of the strengths and limitations of such a tool, we argue for a new small/medium space mission dedicated to gathering high-precision, high-cadence, long photometric series in dense stellar fields. These crowded fields are being avoided in all present and planned exoplanet searches, which instead favour large pixel sizes and nearby bright stars.

The proposed science case / mission will lead to breakthroughs in stellar astrophysics, especially in the metal poor regime, by performing high-precision tests of stellar physics  in open and, crucially, globular clusters (see Sec. \ref{sec:SO1}). 
Moreover, precise and accurate asteroseismic constraints on global stellar properties, together with complementary astrometric and spectroscopic data, will address unanswered, long-standing questions about the evolution and  formation of open and globular clusters (see Sec. \ref{sec:clusterform}), and aid our understanding 
of the assembly history and chemical evolution of the Milky Way's bulge and few nearby dwarf galaxies (see Sec. \ref{sec:SO3}).

We use the knowledge acquired from the analysis and interpretation of  existing seismic data to propose a possible mission design, including a list of potential targets. The details of what can be achieved depend on the telescope collecting area, on the length and cadence of observations, and the distance of the selected targets. With a few reasonable assumptions, we identify the following priorities, which we show can be reached with a $\sim90$cm-wide telescope with a $\sim1^\circ$ field-of-view, and a combination of 24-, 12-, and 6-months long observing runs on the targets identified in Table \ref{tab:targets}: 
\begin{enumerate}
\item to detect solar-like oscillations in stars with luminosities lower than the red clump, enabling precise and accurate characterisation of  global properties (masses, radii, distances, ages) for stars belonging to a few globular clusters of varying metallicity, open clusters of varying age, and building blocks of the Milky Way;
\item in  globular clusters, calibration of poorly understood processes of stellar evolution, like diffusive processes, mass loss on the red-giant branch, characterisation of horizontal branches and identification of the physical mechanisms determining their morphologies, evolution, and occurrence of of interacting binaries;-- in general, calibration of stellar models at low metallicities; in addition, direct measurements of their helium content and ages, tight constraints to dynamical models, and to the origin of their multiple populations; 
\item in the open clusters, characterization of convective mixing zones and core growth  across the HR diagram, distributions and evolution of rotational velocities, orientation of spin axes, transport of elements and angular momentum; in M67, detailed asteroseismology of solar twins and their parents/progeny; -- in general, the ultimate data for calibrating stellar models at near-solar metallicities;
\item to characterize the stellar populations and their age distributions in the core of the nearby dwarf galaxies $\omega$~Cen and Sgr~dSph, and in the Milky Way's bulge. 
\end{enumerate}
Moreover, the high-precision, high-cadence, long-duration photometric data gathered by such a  mission will also enable us to advance our understanding of stellar activity, flares,  products of binary interactions, and  to constrain the occurrence and properties of exoplanets in different (and largely unexplored) environments, widening the impact of the proposed science case even further.

\begin{acknowledgements}
AM, JM and FV acknowledge support from the European Research Council Consolidator Grant funding scheme (project {ASTEROCHRONOMETRY}, G.A. n. 772293, \url{http://www.asterochronometry.eu}). AM, BM and LG are grateful to the International Space Science Institute (ISSI) for support provided to the \mbox{asteroSTEP} ISSI International Team. AM and WJC acknowledge the support of the UK Science and Technology Facilities Council (STFC).
%\Gael{SNF: 
GB is sponsored by the Swiss National Science Foundation (project number $200020-172505$).
%}
WC acknowledges funding from the Swiss National Science Foundation under grant P400P2\_183846.
We thank G. R. Davies for providing the power spectra used in Fig. \ref{fig:PS}. N.L. acknowledges financial support from "Programme
National de Physique Stellaire" (PNPS) of CNRS/INSU, France.
CC acknowledges partial support from DFG Grant CH1188/2-1 and from the ChETEC COST Action (CA16117), supported by COST (European Cooperation in Science and Technology).
\end{acknowledgements}


\bibliographystyle{spbasic}      % basic style, author-year citations
\bibliography{references}   % name your BibTeX data base

%% Non-BibTeX users please use
%\begin{thebibliography}{}
%%
%% and use \bibitem to create references. Consult the Instructions
%% for authors for reference list style.
%\end{thebibliography}

\end{document}